\documentclass[aps,prl,amsmath,amssymb,floatfix,twocolumn,amsmath,superscriptaddress,twocolumn,nofootinbib,tighten,letterpaper]{revtex4}
\usepackage{multirow}
\usepackage{bbold}
\usepackage{subfigure}
\usepackage{color}
\usepackage{mathrsfs}
\usepackage{hyperref}
\usepackage[normalem]{ulem}
\usepackage{bm}

\usepackage{amsfonts, relsize, color}
\usepackage{graphics}
\usepackage{graphicx}
\usepackage{subfigure}
\usepackage{hyperref}
\usepackage{color}

\begin{document}
\title{\bf Antiunitary symmetry protected higher order topological phases}

\author{Bitan Roy}~\email{bir218@lehigh.edu}
\affiliation{Department of Physics, Lehigh University, Bethlehem, Pennsylvania, 18015, USA}
\affiliation{Max-Planck-Institut f\"{u}r Physik komplexer Systeme, N\"{o}thnitzer Str. 38, 01187 Dresden, Germany}

\date{\today}
\begin{abstract}
Higher-order topological (HOT) phases feature boundary (such as corner and hinge) modes of codimension $d_c>1$. We here identify an \emph{antiunitary} operator that ensures the spectral symmetry of a two-dimensional HOT insulator and the existence of cornered localized states ($d_c=2$) at precise zero energy. Such an antiunitary symmetry allows us to construct a generalized HOT insulator that continues to host corner modes even in the presence of a \emph{weak} anomalous Hall insulator and a spin-orbital density wave orderings, and is characterized by a quantized quadrupolar moment $Q_{xy}=0.5$. Similar conclusions can be drawn for the time-reversal symmetry breaking HOT $p+id$ superconductor and the corner localized Majorana zero modes survive even in the presence of weak Zeeman coupling and $s$-wave pairing. Such HOT insulators also serve as the building blocks of three-dimensional second-order Weyl semimetals, supporting one-dimensional hinge modes.             
\end{abstract}

\maketitle

\begin{figure}[t!]
\includegraphics[height=6.5cm,width=7.25cm]{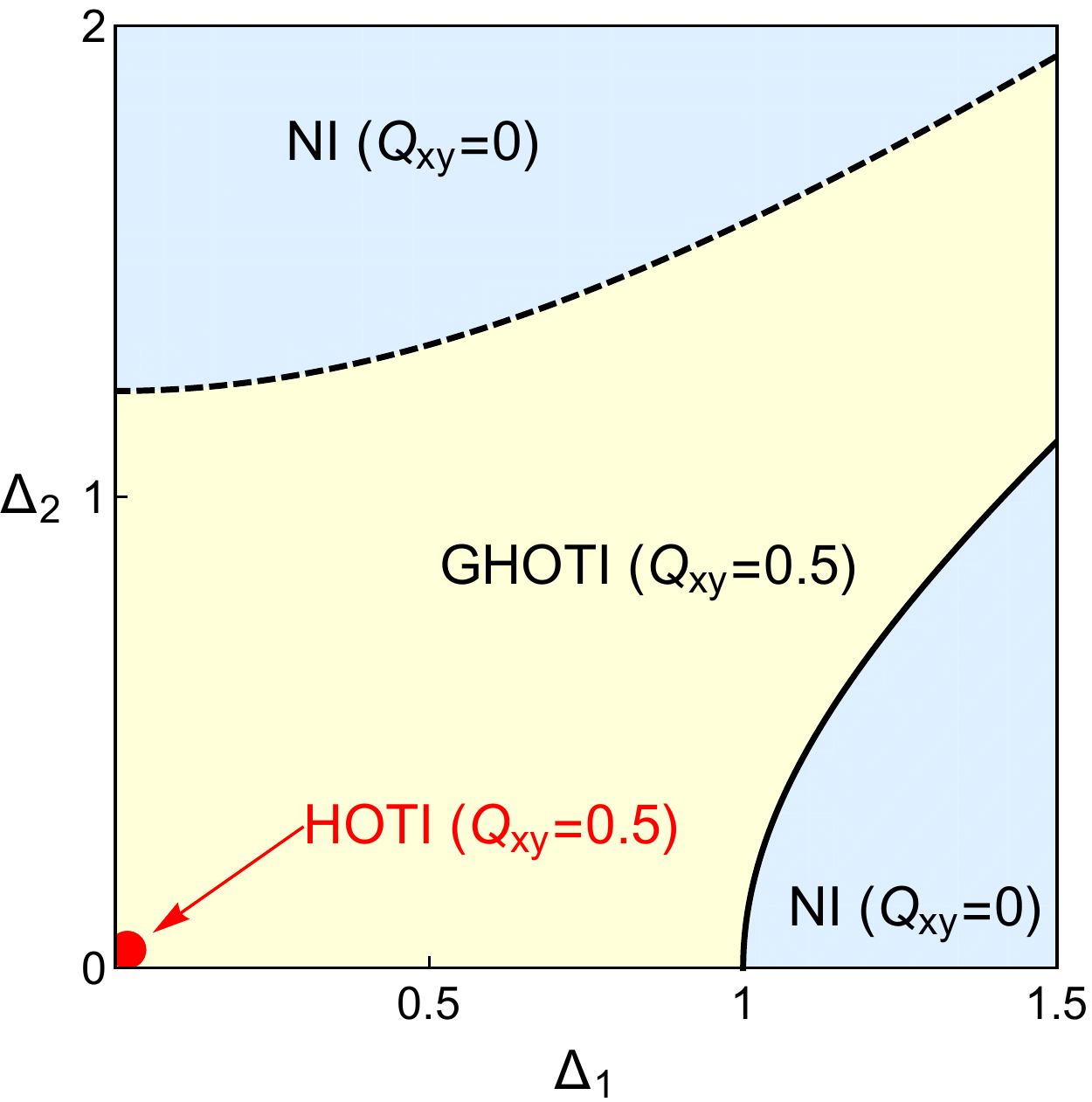}
\caption{Phase diagram of a time-reversal and $C_4$ symmetry breaking Dirac insulator, represented by the Hamiltonian operator $H^{\rm gen}_{\rm HOT}=\hat{h}_0+ \hat{h}_1 + \hat{h}_{\rm p}$ for $t=t_0=m=1$ [Eqs.~(\ref{Eq:HOTImodel}) and ~(\ref{Eq:genHOTIpert})]. For small $\Delta_1$ and $\Delta_2$, the system supports four zero-energy corner modes (Fig.~\ref{Fig:GHOTI_CornerModes}), protected by an antiunitary operator ($A$) and representing a generalized higher order topological insulator (GHOTI). For charged fermions GHOTI is characterized by a quantized quadrupolar moment $Q_{xy}=0.5$. The bulk band gap closes either at the $\Gamma$ point (solid line) or along the $\Gamma-{\rm M}$ line (dashed line) (Fig.~\ref{Fig:bandstructure}) beyond which the system becomes a trivial or normal insulator (NI), with $Q_{xy}=0$ for charged fermions. The phase diagram possesses a \emph{reflection} symmetry about $(\Delta_1,\Delta_2)=(0,0)$, where the bands recover two-fold degeneracy [see Fig.~\ref{Fig:bandstructure} (left column)], and the system describes a regular HOTI (red dot). The phase boundaries do not depend on $\Delta$ ($C_4$ symmetry breaking mass).      
}~\label{Fig:PD_bandstructure}
\end{figure}

\emph{Introduction}. The hallmark of topological phases of matter is the presence of gapless modes at the boundary, protected by the nontrivial bulk topological invariant. Traditionally, a $d$-dimensional bulk topological phase (insulating or gapless) hosts boundary modes that are localized on $d-1$ dimensional surfaces, characterized by codimension $d_c=1$~\cite{hasan-kane-review2010, qi-zhang-review2011, bernevig2006, fu-kane2007,volovik-book, chiu-review2016, armitage-review2018}. Nevertheless, the family of topological phases of matter nowadays includes its higher order cousins, and an $n$th-order topological phase features boundary modes of codimension $d_c=n>1$~\cite{benalcazar2017}, such as the corner (with $d_c=d$) and hinge (with $d_c=d-1$) states of topological insulators (electrical and thermal) and semimetals~\cite{benalcazar2017, schindler2018, serra-garcia2018, noh2018, peterson,imhof2018, song2017, langbehn2017, franca2018,schindler-sciadv2018, ezawa2018, hsu2018, lin-hughes-DSM, wang1-2018, yan2018, calugaru2019, okuma2018, tanay2019, sigrist2019, benalcazar-prb2017, matsugatani2018, khalaf2018, Vliu2018, seradjeh2018, ahn2018, Klinovaja2019, Klinovaja2019arXiv, kaisun2019arXiv, ghorashi-HOTSC, vanmiert2018, wang-arxiv2018, trifunovic2019}. In this language, the traditional topological phases are first order. While the bulk topological invariant assures the existence of boundary modes, often (if not always) the localized topological modes get pinned at precise zero energy due to the \emph{spectral symmetry}, which we exploit here to propose the most general setup for a two-dimensional higher-order topological (HOT) insulator, characterized by a quantized quadrupolar moment $Q_{xy}=0.5$ and supports four corner localized zero-energy modes. The central results are summarized in the phase diagram, shown in Fig.~\ref{Fig:PD_bandstructure}.

\begin{figure*}[t!]
\includegraphics[width=0.24\linewidth]{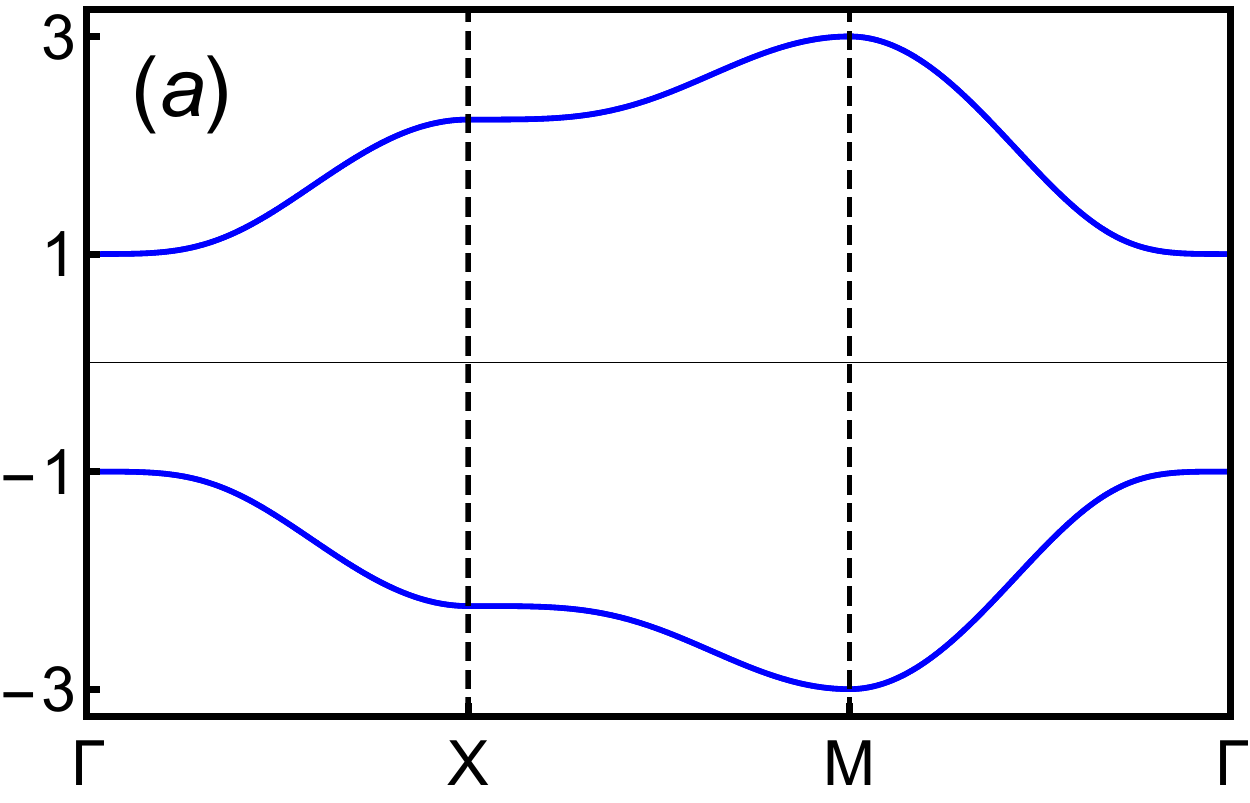}
\includegraphics[width=0.24\linewidth]{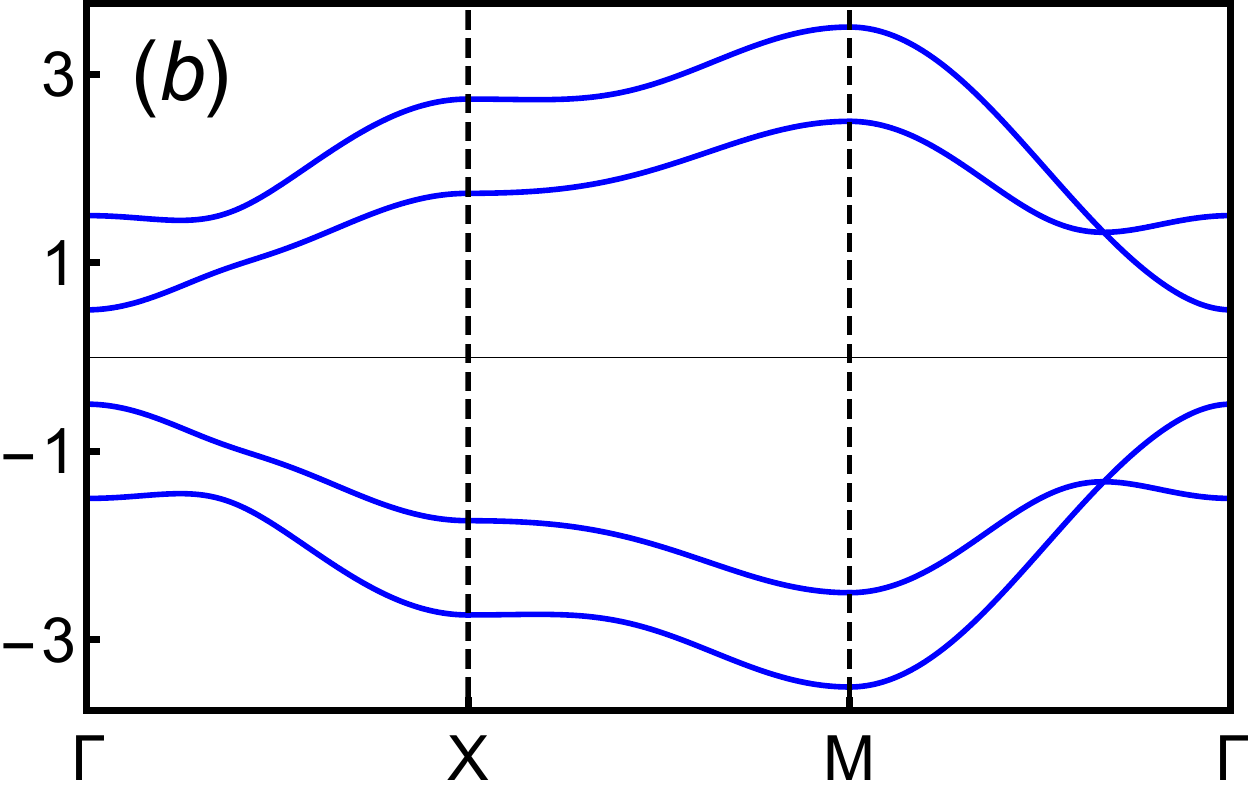}
\includegraphics[width=0.24\linewidth]{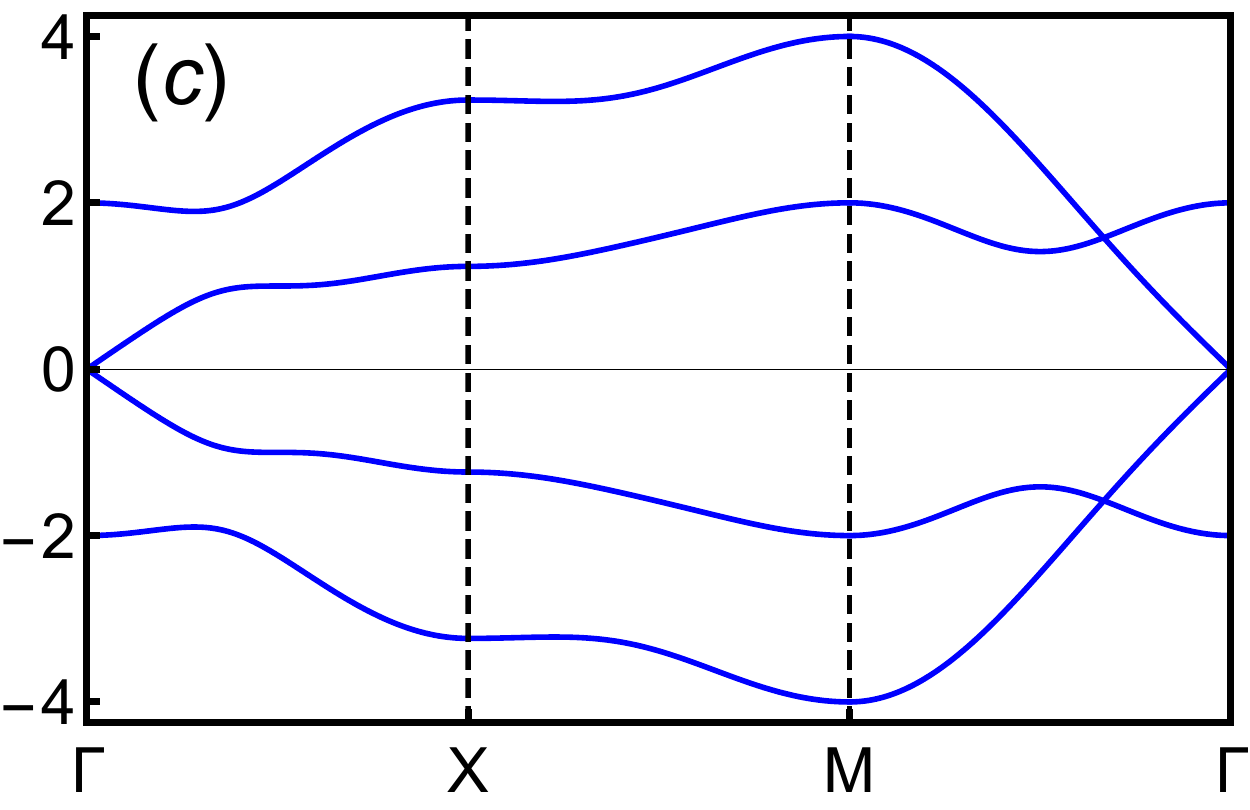}
\includegraphics[width=0.24\linewidth]{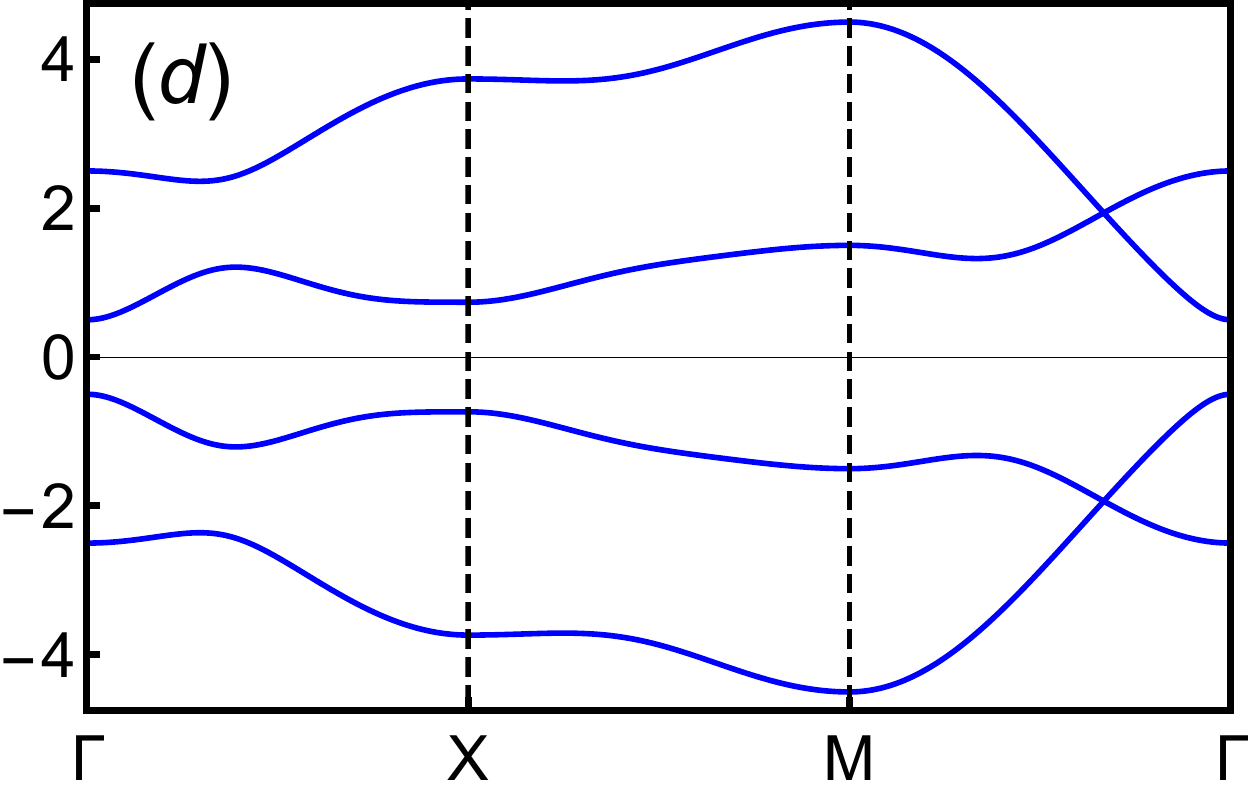}
\includegraphics[width=0.24\linewidth]{HOTI_BandStructure_NoPert.pdf}
\includegraphics[width=0.24\linewidth]{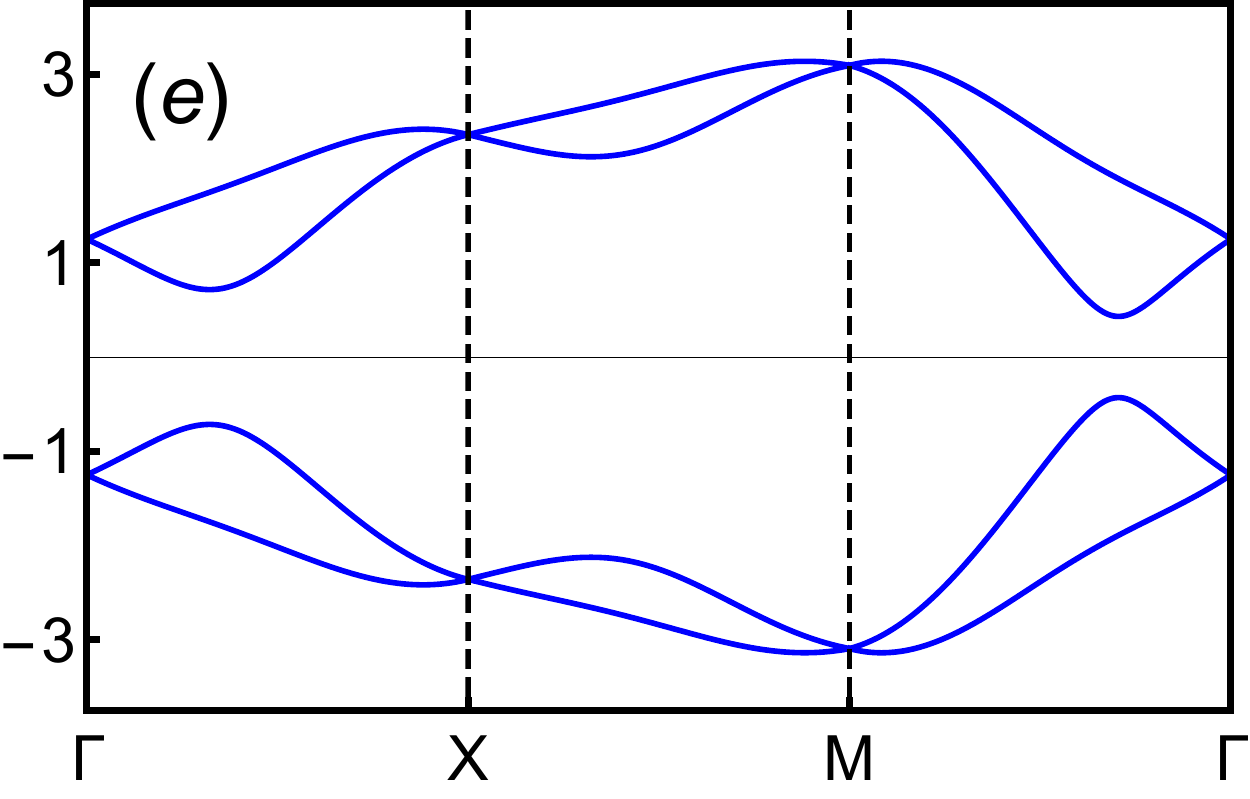}
\includegraphics[width=0.24\linewidth]{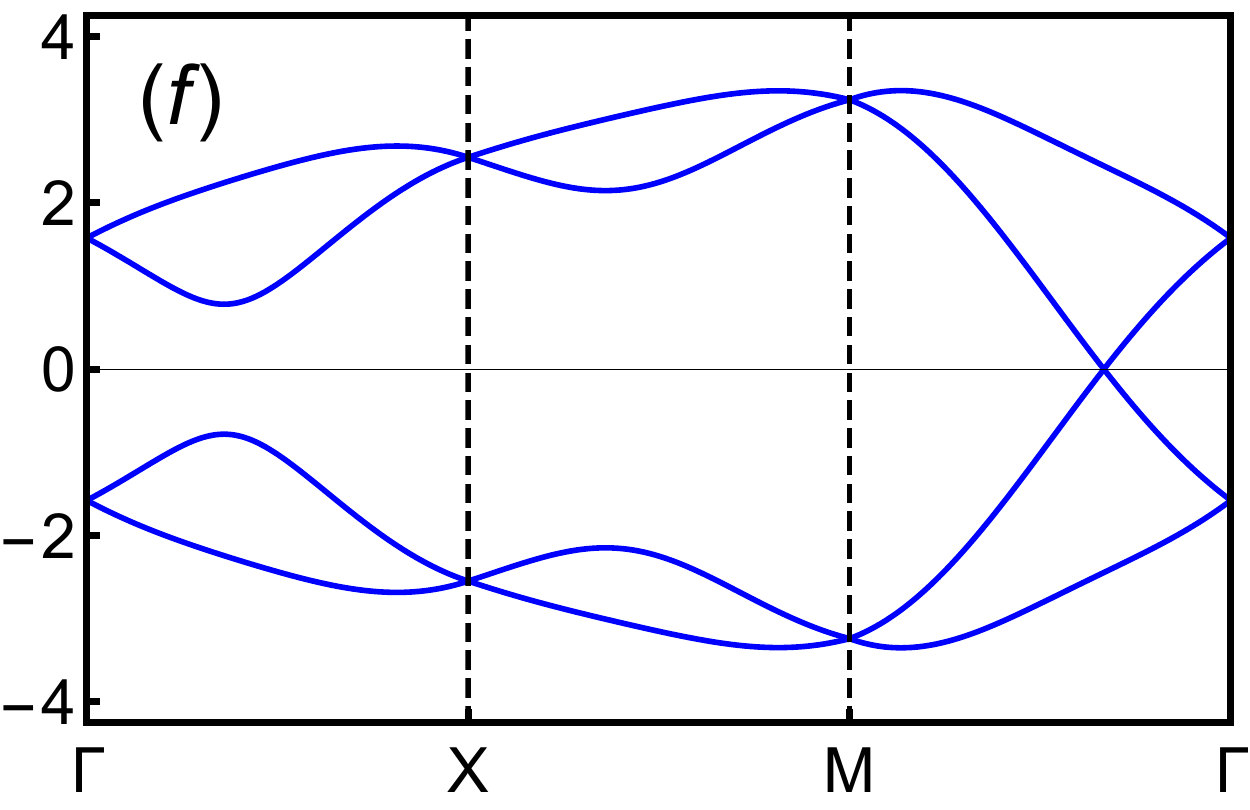}
\includegraphics[width=0.24\linewidth]{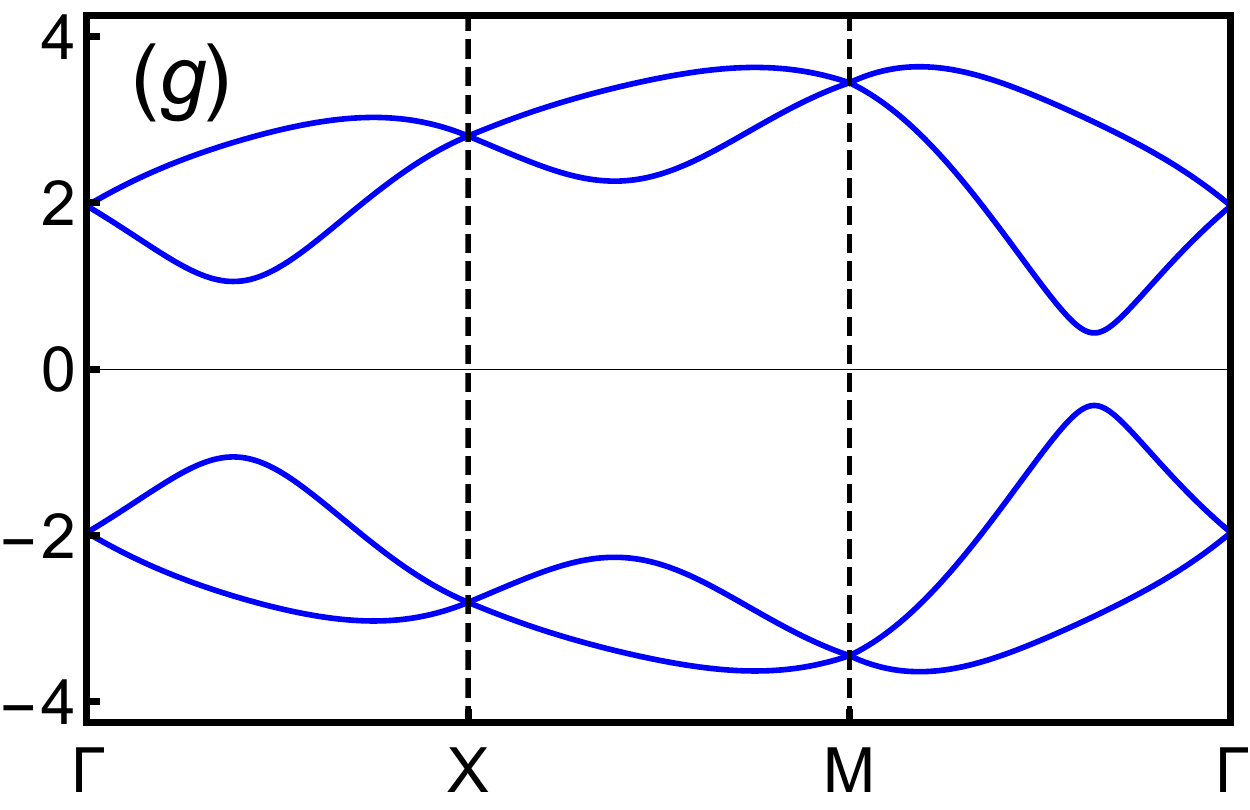}
\caption{(a) Band structure of a time-reversal and four-fold symmetry breaking two-dimensional Dirac insulator (bands possess two-fold degeneracy), described by the tight-binding model from Eq.~(\ref{Eq:HOTImodel}) for $t=t_0=m=\Delta=1$, and $\Delta_1=\Delta_2=0$. Along the vertical axis energy is measured in units of $t$. Effects of two perturbations [$\Delta_1$ and $\Delta_2$, see Eq.~(\ref{Eq:genHOTIpert})] are shown in the rest of the panels. We set $\Delta_1=0.50$ in (b), $1.00$ in (c) and $1.50$ in (d), and $\Delta_2=0$ for (b)-(d). For (e)-(g) $\Delta_1=0$, but $\Delta_2=0.75$ in (e), $1.22$ in (f) and $1.69$ in (g). We here take the path $\Gamma \to {\rm X} \to {\rm M} \to \Gamma$ in the Brillouin zone, where $\Gamma=(0,0)$, ${\rm X}=(0,1)\frac{\pi}{a}$ and ${\rm M}=(1,1)\frac{\pi}{a}$, and $a$ is the lattice spacing. The band gap closes either at the $\Gamma$ point [see (c)] or along the $\Gamma-{\rm M}$ line [see (f)], respectively yielding the phase boundaries between GHOTI and NI, shown by the solid and dashed lines in Fig.~\ref{Fig:PD_bandstructure}. 
}~\label{Fig:bandstructure}
\end{figure*}

The HOT phases can be constructed (at least, in principle) by systematically reducing the dimensionality of the boundary modes at the cost of some discrete crystalline and fundamental (such as time-reversal) symmetries in the bulk of the system. For example, a two-dimensional HOT insulator, supporting four corner localized zero-energy modes ($d=0$, $d_c=2$), can be realized in the presence of a four-fold ($C_4$) and time-reversal (${\mathcal T}$) symmetry breaking perturbation that acts as a mass for two one-dimensional counter propagating helical edge modes ($d=1$, $d_c=1$) of a first-order topological insulator. The corresponding effective single-particle Hamiltonian can be decomposed as $\hat{h}^{\rm 2D}_{\rm HOT}=\hat{h}_{0}+ \hat{h}_{1}$, with
\allowdisplaybreaks[4]
\begin{eqnarray}~\label{Eq:HOTImodel}
\hat{h}_{0}  &=& t \sum^2_{j=1} \sin(k_j a) \Gamma_j + \big[ m + t_0 \sum^{2}_{j=1} \left( \cos (k_j a)-1 \right) \big] \Gamma_3, \nonumber \\
\hat{h}_{1} &=& \Delta \left[ \cos(k_x a) - \cos(k_y a) \right] \Gamma_4,
\end{eqnarray}
where $\Gamma_j$'s are mutually anticommuting four-component Hermitian matrices, satisfying $\{ \Gamma_j, \Gamma_k \}=2 \delta_{jk}$ for $j,k=1, \cdots, 5$, $a$ is the lattice spacing (set to be unity) and ${\bf k}$ is spatial momenta. For $0<m/t_0<2$, $\hat{h}_{0}$ describes a first-order topological insulator. But, depending on the spinor basis and the corresponding representation of $\Gamma$ matrices (about which more in a moment), this phase represents a quantum spin-Hall insulator (QSHI) or a topological $p$-wave pairing. On the other hand, $\hat{h}_1$ lacks both $C_4$ and ${\mathcal T}$ symmetries. It (1) acts as a mass for the edge modes, since $\{ \hat{h}_1, \hat{h}_0 \}=0$, and (2) changes sign under the $C_4$ rotation, thus assuming the profile of a domain wall. Then a generalized Jackiw-Rebbi index theorem~\cite{Jackiw-Rebbi}, guarantees the existence of four corner localized zero energy modes, with $d_c=2$. We then realize a second-order topological insulator. Respectively for charged and neutral fermions, $\hat{h}_1$ represents either a spin-orbital density wave ordering and a $d$-wave pairing. In the latter case, the resulting phase stands as HOT $p+id$ pairing~\cite{wang1-2018}.

\begin{figure*}[t!]
\includegraphics[width=0.19\linewidth]{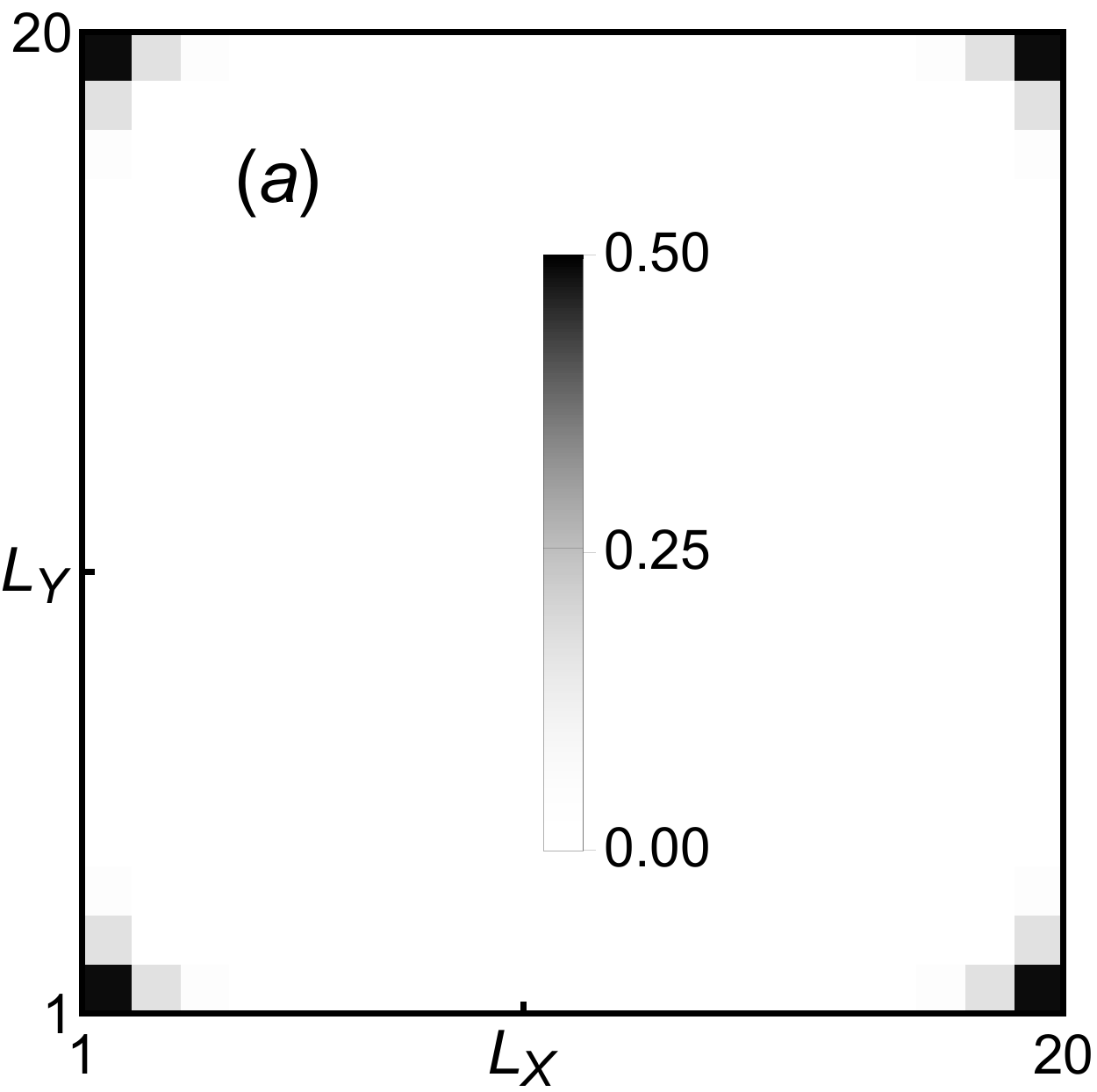}%
\includegraphics[width=0.19\linewidth]{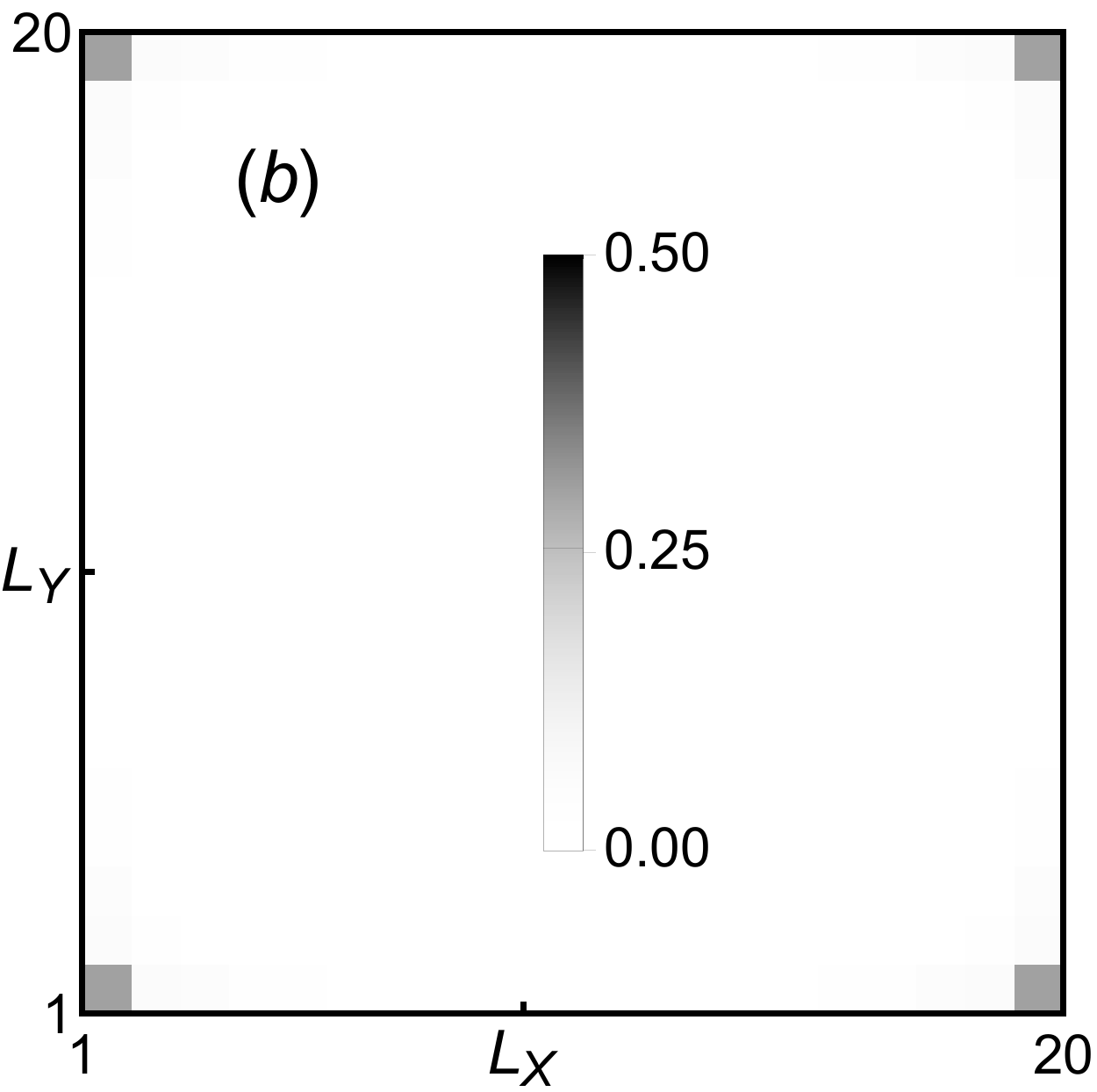}%
\includegraphics[width=0.19\linewidth]{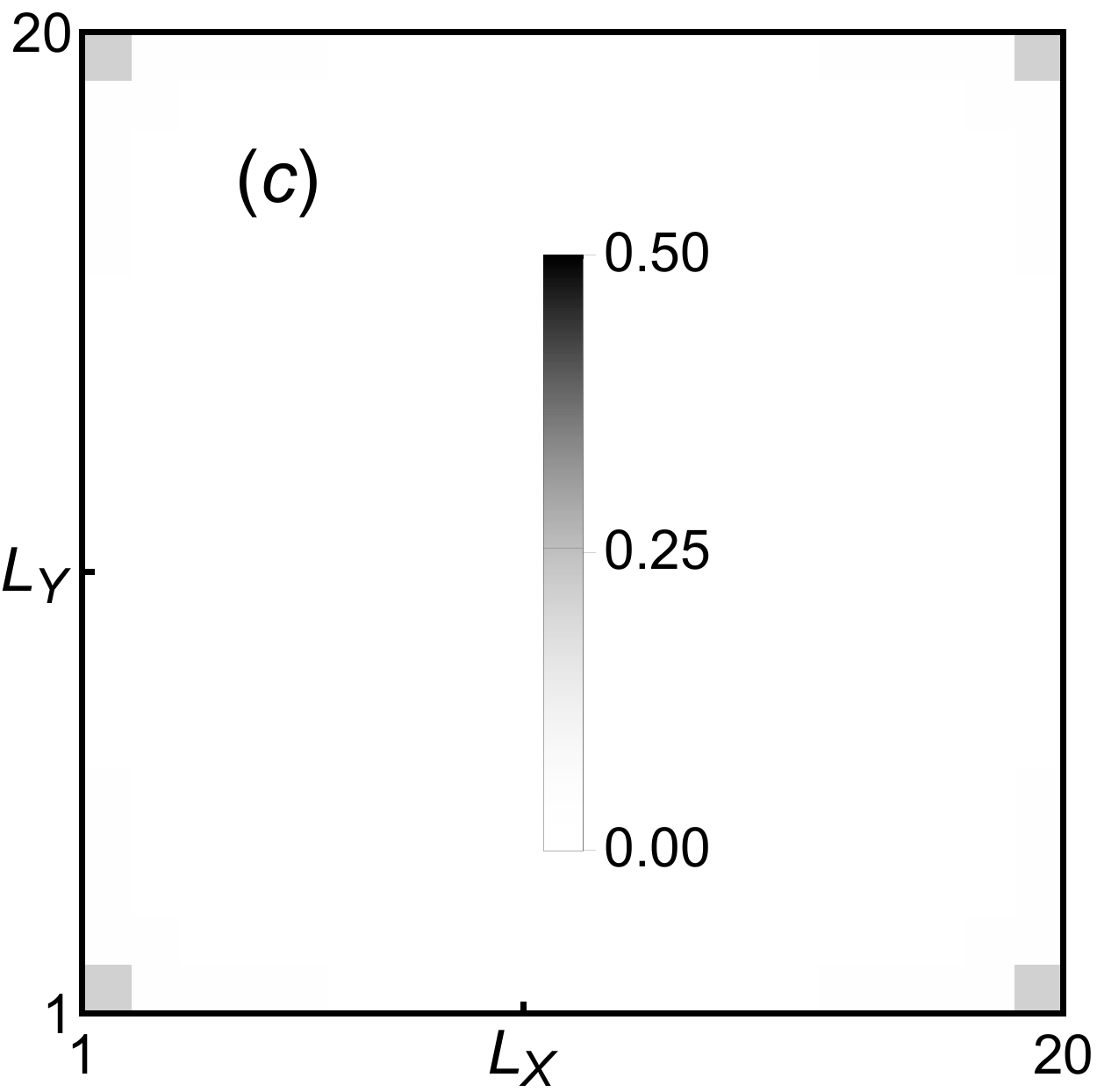}%
\includegraphics[width=0.19\linewidth]{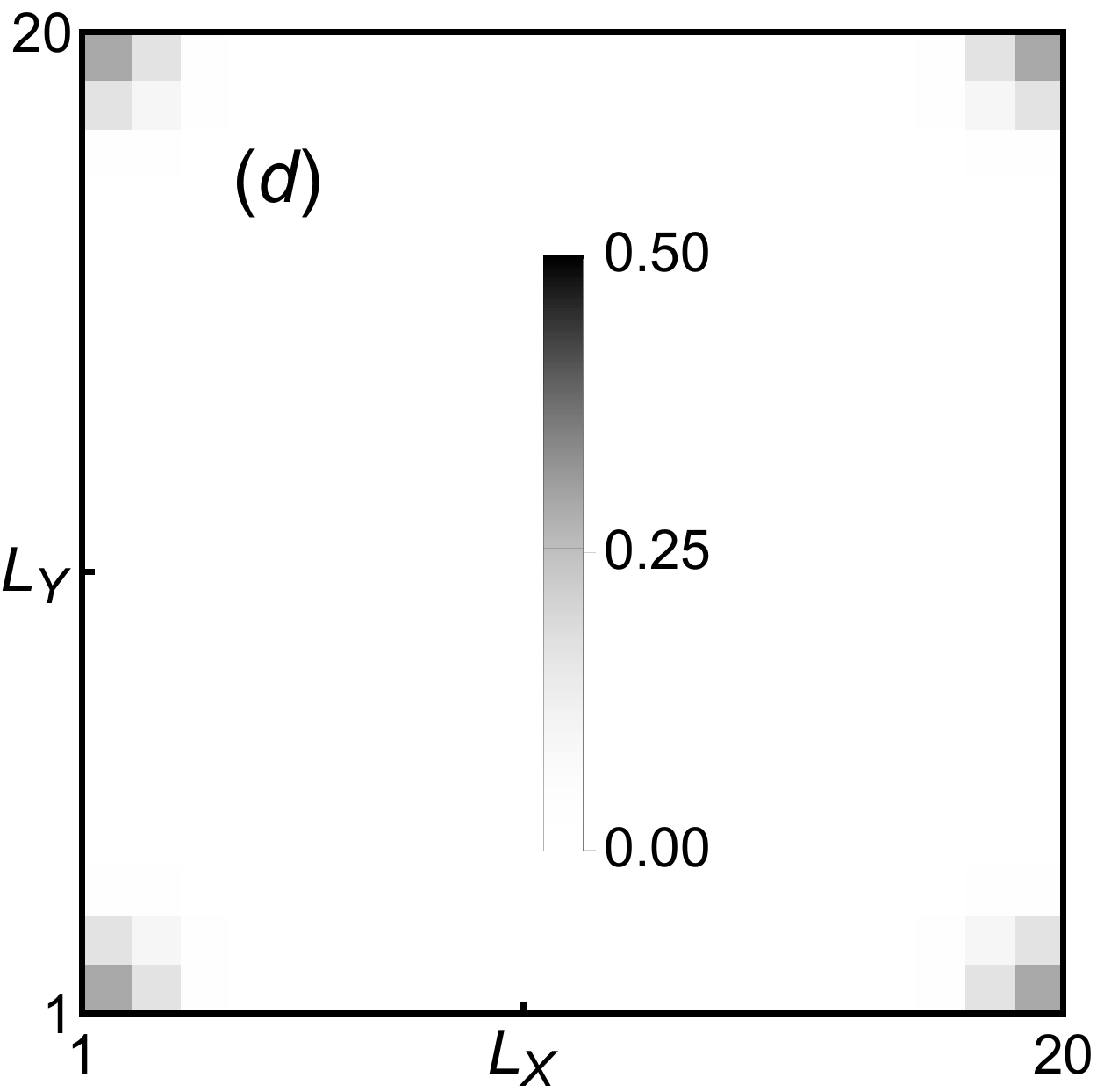}%
\includegraphics[width=0.19\linewidth]{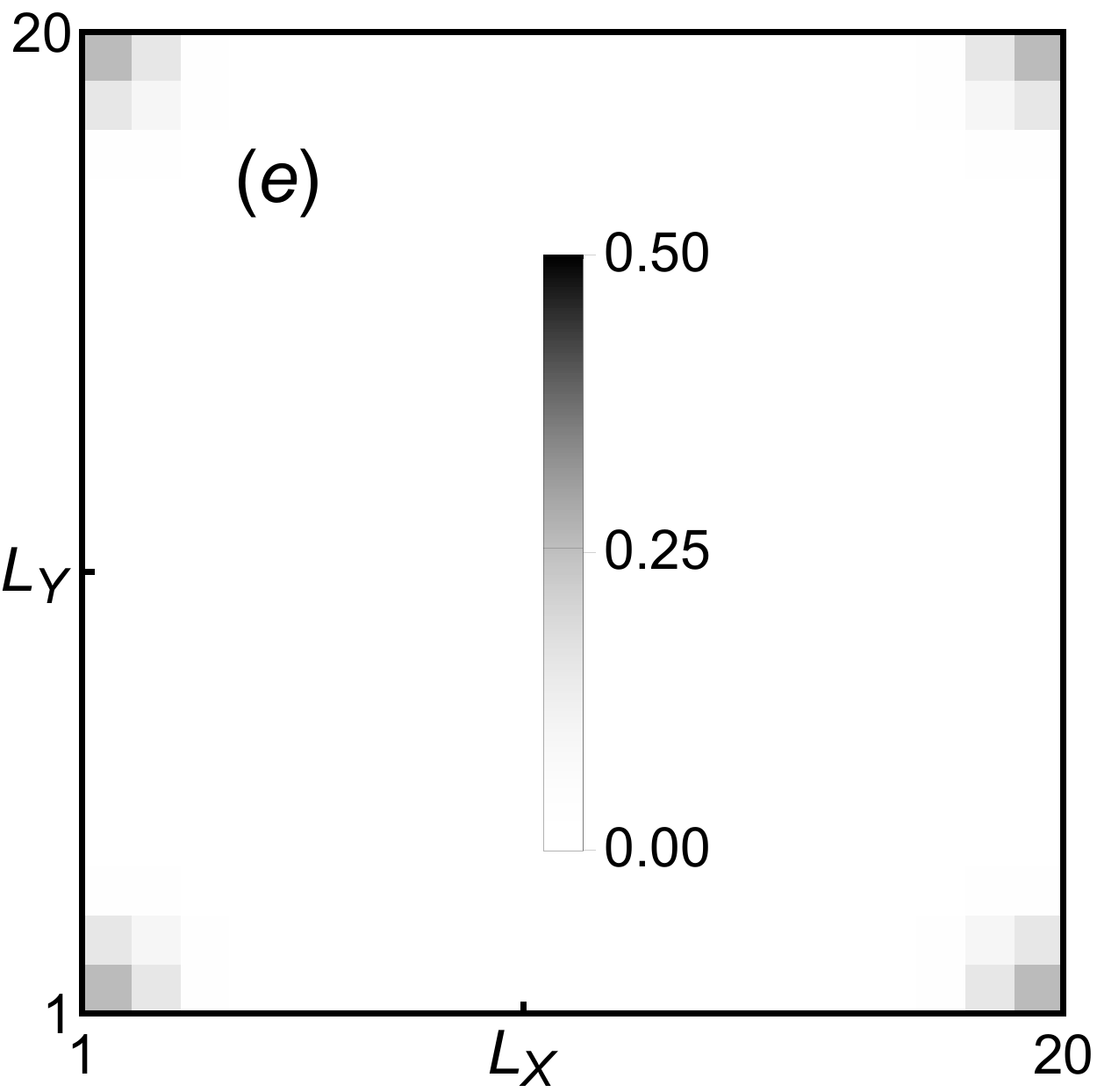}
\caption{Local density of states associated with four near (due to finite system size) zero-energy corner modes for (a) $\Delta_1=0=\Delta_2$ (regular HOT insulator), (b) $\Delta_1=0.50, \Delta_2=0$, (c) $\Delta_1=0.85,\Delta_2=0$, (d) $\Delta_1=0,\Delta_2=0.50$ and (e) $\Delta_1=0,\Delta_2=0.80$ [see Eqs.~(\ref{Eq:HOTImodel}) and ~(\ref{Eq:genHOTIpert})]. With increasing $\Delta_1$ or $\Delta_2$, even though the corner localization of zero modes decreases monotonically, the system continues to describe a GHOTI with quantized quadrupole moment $Q_{xy}=0.5$, when they are \emph{small} (see Fig.~\ref{Fig:PD_bandstructure}). Also in a periodic system (devoid of a corner mode) $Q_{xy}=0.5$, suggesting that GHOTI is a bulk topological phase. For large enough $\Delta_1$ or $\Delta_2$, the corner modes disappear (not shown explicitly) and system becomes a trivial insulator. Numerical analyses are performed in a system with linear dimension $L=20$ in both the $x$ and $y$ directions, and for $t=t_0=m=1$ and $\Delta=0.50$.     
}~\label{Fig:GHOTI_CornerModes}
\end{figure*}

We here seek to answer the following question. \emph{What is the most general form of the Hamiltonian operator that supports topologically protected corner modes at precise zero energy and describes a two-dimensional HOT insulator?} We note that the corner modes are pinned at zero energy due to the spectral symmetry of $\hat{h}^{\rm 2D}_{\rm HOT}$, generated by a unitary ($U$) as well as an antiunitary ($A$) operators, such that $\{ \hat{h}^{\rm 2D}_{\rm HOT}, U \}=0=\{ \hat{h}^{\rm 2D}_{\rm HOT}, A \}$. Since the maximal number of mutually anticommuting four-component $\Gamma$ matrices is \emph{five} and only four of them appear in $\hat{h}^{\rm 2D}_{\rm HOT}$, one is always guaranteed to find $U=\Gamma_5$. On the other hand, the existence of $A$ can be assured in the following way. Note all representations of mutually anticommuting four-component Hermitian $\Gamma$ matrices are \emph{unitarily equivalent}. Hence, without any loss of generality, we commit to a representation in which $\Gamma_1$ and $\Gamma_2$ ($\Gamma_3$ and $\Gamma_4$) are purely real (imaginary)~\cite{okubo,herbut-lu,roy-herbut-halfvortex}. Then $A=K$, where $K$ is the complex conjugation~\cite{antiunitary-comment}. Identification of the antiunitary operator $A$ allows us to construct the most general form of the Hamitonian operator $\hat{h}^{\rm gen}_{\rm HOT}=\hat{h}^{\rm 2D}_{\rm HOT} + \hat{h}_p$, such that $\{ \hat{h}^{\rm gen}_{\rm HOT}, A \}=0$ (with real $\Delta_1$ and $\Delta_2$), where 
\begin{equation}~\label{Eq:genHOTIpert}
\hat{h}_{\rm p}=  \Delta_1 \; (i \Gamma_1 \Gamma_2) +  \Delta_2 \; (i \Gamma_3 \Gamma_4)
\equiv \Delta_1 \; \Gamma_{12} + \Delta_2 \; \Gamma_{34}. 
\end{equation}
For \emph{small} $\Delta_1$ and $\Delta_2$, the system continues to support four zero-energy corner modes [see Fig.~\ref{Fig:GHOTI_CornerModes}] and a quantized quadrupolar moment $Q_{xy}=0.5$ (modulo 1). The resulting phase then describes a two-dimensional \emph{generalized} higher order topological insulator (GHOTI). However, for sufficiently large $\Delta_1$ or $\Delta_2$, the system enters into a trivial or normal insulating phase, where $Q_{xy}=0$ (modulo 1), following a band gap closing (see Fig.~\ref{Fig:bandstructure}). These findings are summarized in Fig.~\ref{Fig:PD_bandstructure}. The physical meanings of $\Delta_1$ and $\Delta_2$ are of course representation dependent~\cite{antiunitary-wavefunction}.

\emph{Charged fermions}. We first focus on charged fermions and introduce a four-component spinor $\Psi^\top_{\bf k}= \big( c^{\bf k}_{A,\uparrow}, c^{\bf k}_{B,\uparrow}, c^{\bf k}_{A,\downarrow}, c^{\bf k}_{B,\downarrow} \big)$, where $c^{\bf k}_{X,\sigma}$ is the fermion annihilation operator on sublattice/orbital $X=A,B$ with spin projection $\sigma=\uparrow, \downarrow$ and momenta ${\bf k}$. Then $\hat{h}_0$ describes a QSHI (for $0<m/t_0<2$), when the $\Gamma$ matrices are $\Gamma_1=\sigma_3 \tau_1$, $\Gamma_2=\sigma_0 \tau_2$, $\Gamma_3=\sigma_0 \tau_3$, $\Gamma_4=\sigma_1 \tau_1$ and $\Gamma_5=\sigma_2 \tau_1$. The Pauli matrices ${\boldsymbol \sigma} ({\boldsymbol \tau})$ operate on the spin (sublattice/orbital) degrees of freedom. In this representation $A=\Gamma_1 K$, and $\Delta_1 \; (\Delta_2)$ corresponds to anomalous charge Hall (spin and orbital density-wave) order.

Note that $\hat{h}_0$ preserves both time-reversal (${\mathcal T}$) and parity (${\mathcal P}$) or inversion symmetries. Under the reversal of time ${\bf k} \to -{\bf k}$ and $\Psi_{\bf k} \to \sigma_2 \tau_0 \Psi_{-{\bf k}}$. Hence, ${\mathcal T}= \Gamma_1 \Gamma_4 K$ and ${\mathcal T}^2=-1$. Under the spatial inversion ${\bf r} \to -{\bf r}$ and $\Psi_{\bf k} \to \sigma_0 \tau_3 \Psi_{-{\bf k}}$, yielding ${\mathcal P}=\Gamma_3$. By contrast, $\hat{h}_1$ breaks ${\mathcal T}$, ${\mathcal P}$ as well as discrete $C_4$ rotation about the $z$-axis ($\hat{C}^z_4$), under which $(k_x,k_y) \to (k_y,-k_x)$ and $\hat{C}^z_4=\exp[i \frac{\pi}{4} \sigma_3 \tau_3] \equiv \exp[i \frac{\pi}{4} \Gamma_{12}]$. Nonetheless, one can define a `pseudo' time-reversal operator ${\mathcal T}_{\rm ps}=i \sigma_2 \tau_3 K = \Gamma_2 \Gamma_5 K$, under which ${\bf r} \to -{\bf r}$ as well, such that $\left[ \hat{h}^{\rm 2D}_{\rm HOT}, {\mathcal T}_{\rm ps} \right]=0$ and ${\mathcal T}^2_{\rm ps}=-1$. Consequently, the valence and conduction bands of a HOT insulator ($\Delta_1=0=\Delta_2$) possess two-fold degeneracy [see Fig.~\ref{Fig:bandstructure} (first column)].

Once we turn on $\hat{h}_{\rm p}$ [see Eq.~(\ref{Eq:genHOTIpert})], the bands loose the two-fold degeneracy (see Fig.~\ref{Fig:bandstructure}). Note that under ${\mathcal T}$, ${\mathcal P}$ and ${\mathcal T}_{\rm ps}$, the term proportional to $\Delta_1$ ($\Delta_2$) is odd (even), even (odd) and odd (odd). Therefore, it is \emph{impossible} to find an antiunitary operator that commutes with $\hat{h}^{\rm gen}_{\rm HOT}$ and squares to $-1$. As a result, the energy spectra of $\hat{h}^{\rm gen}_{\rm HOT}$ only contains non-degenerate bands. Still $\{ \hat{h}^{\rm gen}_{\rm HOT}, A \}=0$, assuring the spectral symmetry among the bands about the zero energy. It is worth pointing out that $\hat{h}^{\rm gen}_{\rm HOT}$ is \emph{algebraically} similar to the \emph{generalized} Jackiw-Rossi Hamiltonian, yielding zero-energy modes bound to the core of a vortex in $d=2$~\cite{herbut-lu,roy-herbut-halfvortex, Jackiw-Rossi, chamon-GJR, Roy-Goswami-GJR}.

Next, we assess the stability of the HOT insulator in the presence of two perturbations, $\Delta_1$ and $\Delta_2$. As shown in Fig.~\ref{Fig:bandstructure} (second column) that despite loosing the two-fold degeneracy, the bands are still gapped for small $\Delta_1$ and/or $\Delta_2$. But, at an intermediate $\Delta_1$ or $\Delta_2$ the band gap closes either at the $\Gamma$ point (top row) or along the $\Gamma-{\rm M}$ line (bottom row) of the Brillouin zone [see Fig.~\ref{Fig:bandstructure} (third column)]. The line of the band gap closing at the $\Gamma$ point is given by $\Delta_1 =[m^2 + \Delta^2_2]^{1/2}$ (see the solid line in Fig.~\ref{Fig:PD_bandstructure}). On the other hand, the gap closing along the $\Gamma-{\rm M}$ line takes place at momenta ${\bf k}=(\pm,\pm)k_\ast$ and the corresponding phase boundary (the dashed line in Fig.~\ref{Fig:PD_bandstructure}) is determined by $\Delta_2 =[\Delta^2_1 + 2 t^2_0 \; \sin^2 (k_\ast)]^{1/2}$, where $k_\ast=\cos^{-1}\left( \frac{m-2t_0}{2 t_0}\right)$. At the gap closing points, the system is described in terms of linearly dispersing massless two-component Weyl fermions at low energies. For stronger $\Delta_1$ or $\Delta_2$, the system reenters into an insulating (but trivial) phase (see the fourth column of Fig.~\ref{Fig:bandstructure}). Note that the phase boundaries between GHOTI and the trivial insulator do not depend on $\Delta$, as $\hat{h}_1$ vanishes at the $\Gamma$ point and along the $\Gamma - {\rm M}$ line.

We now anchor the topological nature of these insulators, separated by a band gap closing. To this end, we numerically diagonalize the effective tight-binding model, namely $\hat{h}^{\rm gen}_{\rm HOT}$, on a square lattice of linear dimension $L$ and with an open boundary in each direction for various choices of $\Delta_1$ and $\Delta_2$. The results are shown in Fig.~\ref{Fig:GHOTI_CornerModes}. For $\Delta_1=0=\Delta_2$, the system supports four near (due to a finite system size) zero energy states that are highly localized near the corner of the system, yielding a conventional HOT insulator [see Fig.~\ref{Fig:GHOTI_CornerModes}(a)].

An HOT insulator can be identified from the quantized quadrupolar moment $Q_{xy}=1/2$ (modulo 1)~\cite{multipole1, multipole2, agarwala}. In order to compute $Q_{xy}$, we first evaluate 
\begin{equation}
n={\rm Re} \left[ -\frac{i}{2 \pi} {\rm Tr} \left( \ln \left\{ U^\dagger  \exp \left[ 2 \pi i \sum_{\bf r} \hat{q}_{xy} ({\bf r}) \right]  U \right\} \right) \right],
\end{equation}
where $\hat{q}_{xy} ({\bf r})= x y \; \hat{n}({\bf r})/L^2$ and $\hat{n}({\bf r})$ is the number operator at ${\bf r}=(x,y)$, and $U$ is constructed by columnwise arranging the eigenvectors for the negative energy states. The quadrupolar moment is defined as $Q_{xy}=n-n_{\rm al}$, where $n_{\rm al}=(1/2) \; \sum_{\bf r} x y /L^2$ represents $n$ in the atomic limit and at half filling. Indeed for a HOT insulator, we find $Q_{xy}=0.5$ (within numerical accuracy). While a quantized quadrupolar moment is solely supported by the $C_4$ symmetry breaking Dirac mass ($\hat{h}_1$), the antiunitary operator ($A$) allows us to construct GHOTI.

For finite but small $\Delta_1$ and/ or $\Delta_2$, the system continues to support four corner localized zero-energy modes, and describes a GHOTI [see Figs.~\ref{Fig:GHOTI_CornerModes}(b)-\ref{Fig:GHOTI_CornerModes}(e)], with $Q_{xy}=0.5$. However, with increasing $\Delta_1$ or $\Delta_2$, they gradually loose support at the corners. But, the system still continues to describe a GHOTI up to critical values of $\Delta_1$ and $\Delta_2$. Finally, beyond the band gap closing the system enters into a trivial insulating phase, where $Q_{xy}=0$. Hence, $\hat{h}^{\rm gen}_{\rm HOT}$ describes a HOT phase for small $\Delta_1$ or $\Delta_2$.

Before leaving the territory of charged fermions, we demonstrate the applicability of the above construction of GHOTI in the context of the original model of the two-dimensional HOT insulator introduced in Ref.~\cite{benalcazar2017}, the Belancazar-Bernevig-Hughes (BBH) model. The corresponding Hamiltonian operator reads $\hat{h}^{\rm BBH}_{\rm HOT}=\hat{h}^\prime_1 + \hat{h}^\prime_2$, with
\begin{equation}
\hat{h}^\prime_{j}= \lambda_1 \sin(k_j a) \; \gamma_j + \left[ \beta + \lambda_2 \cos(k_j a) \right] \; \gamma_{2+j},
\end{equation}
for $j=1,2$, where $\gamma_j$'s are mutually anticommuting four-component Hermitian matrices, satisfying $\{ \gamma_j, \gamma_k \}=2\delta_{jk}$. Notice $\hat{h}^\prime_j$ describe Su-Schrieffer-Heeger (SSH) chain in the $x$ and $y$ direction, respectively for $j=1$ and $2$. Specifically for $|\beta/\lambda_2|<1$, each SSH chain supports two endpoint zero energy modes~\cite{SSH-original}. Decoupled $x$ and $y$ SSH chains respectively support a string of such endpoint zero modes along the $y$ and $x$ direction. However, the BBH model supports zero-energy modes only at the four corners, where both SSH chains place endpoint zero modes, yielding a second order topological insulator. This is so, since $\hat{h}^\prime_1$ acts as mass for the zero modes of $\hat{h}^\prime_2$ and vice versa as $\{ \hat{h}^\prime_1, \hat{h}^\prime_2 \}=0$. Notice $\hat{h}^{\rm BBH}_{\rm HOT}$ assumes the form of $\hat{h}^{\rm 2D}_{\rm HOT}$ [see Eq.~(\ref{Eq:HOTImodel})], with $\Gamma_1=\gamma_1$, $\Gamma_2=\gamma_2$, $\Gamma_3=\gamma_+$, $\Gamma_4=\gamma_-$, where $\gamma_\pm =\gamma_3 \pm \gamma_4$, and $t=\lambda_1$, $m=\beta+2t_0$, $t_0=\Delta=\lambda_2/2$. Therefore, our discussion on the GHOTI is equally germane to the BBH model. Without exploiting this correspondence, we can choose (without loss of generality) $\gamma_{1,2}$ ($\gamma_{3,4}$) to be purely real (imaginary), and construct GHOTI from the BBH model, respecting the spectral symmetry generated by $A=K$ and described by the Hamiltonian $\hat{h}^{\rm BBH}_{\rm HOT} + i \Delta_1 \gamma_{1} \gamma_{2} + i \Delta_2 \gamma_{3} \gamma_{4}$.

\emph{HOT pairing}. As a penultimate topic, we focus on two-dimensional HOT superconductor, for which the Nambu spinor is $\Psi^\top_{{\bf k}}= (c_{{\bf k}, \uparrow}, c_{{\bf k}, \downarrow}, c^\ast_{-{\bf k}, \downarrow}, -c^\ast_{-{\bf k}, \uparrow})$ and $c^\ast_{{\bf k},\sigma} (c_{{\bf k},\sigma})$ is the creation (annihilation) operator for the fermionic quasiparticles with momenta ${\bf k}$ and spin projection $\sigma=\uparrow, \downarrow$. The $\Gamma$ matrices are $\Gamma_1=\eta_1 \sigma_1$, $\Gamma_2=\eta_1 \sigma_2$, $\Gamma_3=\eta_3 \sigma_0$, $\Gamma_4=\eta_2 \sigma_0$ and $\Gamma_5=\eta_1 \sigma_3$. The Pauli matrices ${\boldsymbol \eta}$ operate on the Nambu or particle-hole index. The parameter $t$ ($\Delta$) from Eq.~(\ref{Eq:HOTImodel}) represents the amplitude of the $p$ ($d$)-wave pairing, and the term proportional to $t_0$ yields a Fermi surface for $0<m/t_0<2$. Under that circumstance, a weak coupling $p+id$ pairing takes place around the Fermi surface and we realize a second-order topological superconductor, supporting four corner localized Mojorana zero modes~\cite{wang1-2018}. It is worth noting that a mixed parity time-reversal odd $p+is$ pairing, by contrast, only supports gapped Majorana fermions~\cite{goswami-roy-axion}.

In the above-mentioned representation, $\Delta_1$ denotes the Zeeman coupling, while $\Delta_2$ corresponds to the amplitude of spin-singlet (real) $s$-wave pairing. Hence, our discussion on GHOTI suggests that a two-dimensional HOT pairing can be realized in the form of $p+s+id$ pairing even in the presence of (sufficiently weak) Zeeman coupling, at least when the amplitude of the $s$-wave pairing is small enough. Therefore, a quantum phase transition between HOT and a trivial paired state can be triggered by tuning the Zeeman coupling between the quasiparticles and external magnetic field.

\begin{figure}[t!]
\includegraphics[width=0.07\linewidth]{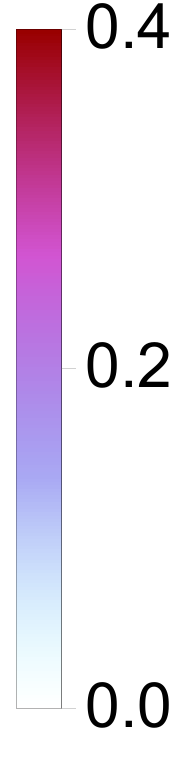}%
\includegraphics[width=0.45\linewidth]{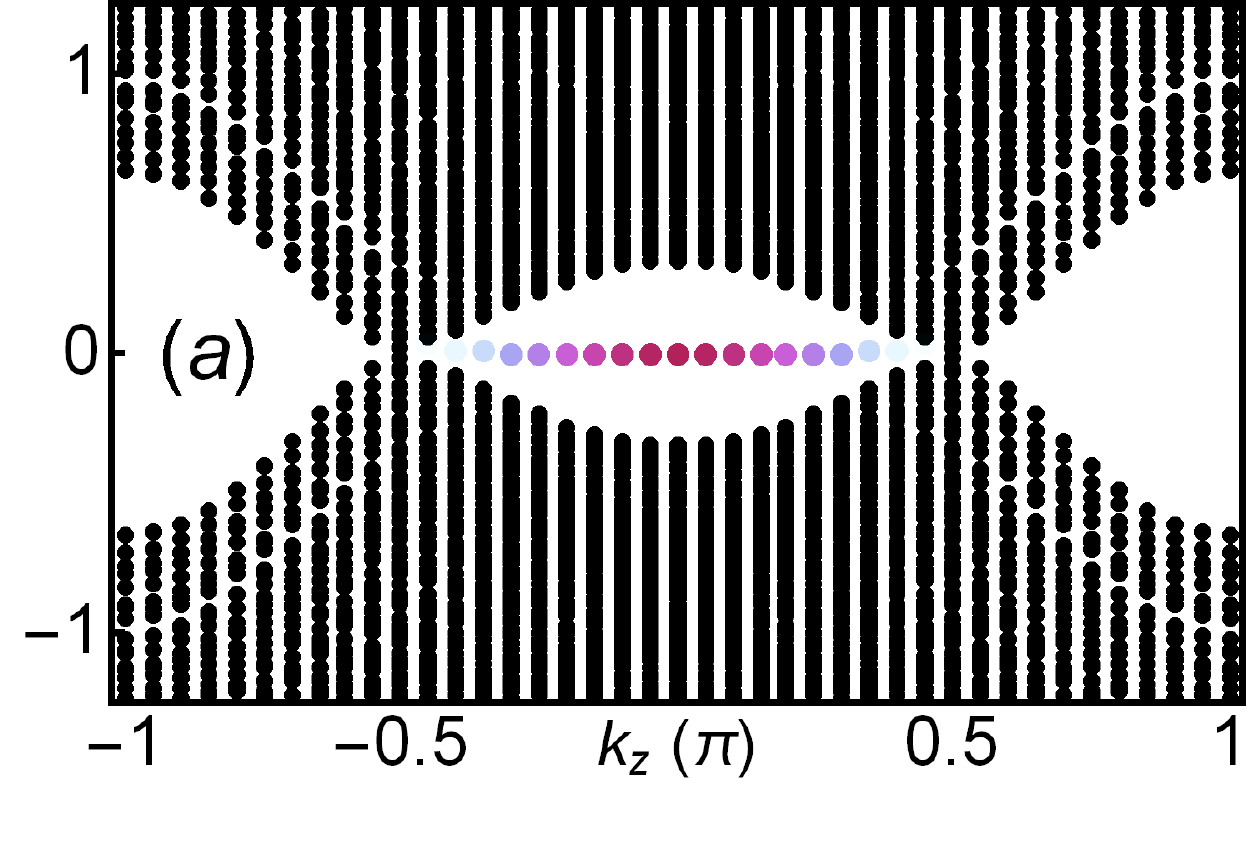}%
\includegraphics[width=0.45\linewidth]{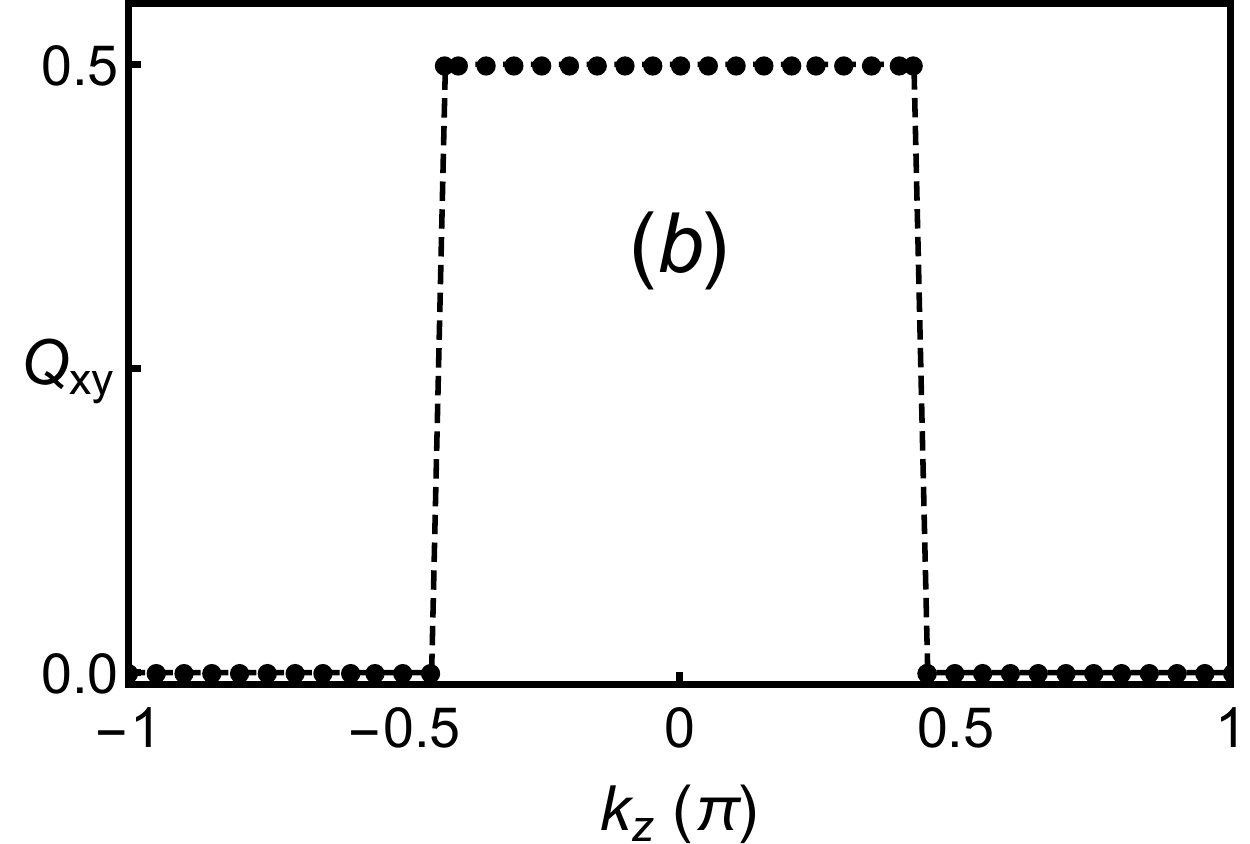}
\caption{(a) Energy spectra (vertical axis) of a second-order Weyl semimetal, supporting one-dimensional zero-energy hinge modes between two Weyl nodes, located at $(0,0,\pm \frac{\pi}{2})$. The corner localization (in color) of the hinge modes decreases monotonically as one approaches the Weyl nodes from the center of the Brillouin zone. (b) Quadrupolar moment ($Q_{xy}$) as a function of $k_z$, showing a quantized value of $0.5$ between two Weyl nodes and $0$ for $|k_z|>\pi/2$. We set $t=t_0=t_z=1$, $m=0$, $\Delta=\Delta_1=\Delta_2=0.5$, and $L_x=L_y=28$.
}~\label{Fig:HOTWeyl}
\end{figure}

Note that when a $d$-wave pairing sets in, it also causes a lattice distortion or electronic nematicity in the system that in turn induces a (small) $s$-wave pairing~\cite{roy-ghorashi-foster-nevidomskyy}. Nonetheless, the amplitude of the $s$-wave pairing can be amplified and the system can also be tuned through the HOT-trivial pairing critical point by applying an external uniaxial strain along the $\langle 11 \rangle$ directions, for example.

\emph{Three dimensions}. Using two-dimensional GHOTI as the building blocks, one can construct three-dimensional HOT phases, by stacking them along the $k_z$ direction in the momentum space. This is accomplished by replacing the term proportional to $\Gamma_3$ in Eq.~(\ref{Eq:HOTImodel}) by
\begin{equation}
\Gamma_3 \left[ t_z \cos(k_z a) + m + t_0 \; \left\{ \cos(k_x a)+\cos(k_y a)-2 \right\} \; \right]. \nonumber 
\end{equation}  
For example, when $\Delta_1=\Delta_2$, the system describes a second order Weyl semimetal (since all bands are non-degenerate) with two Weyl nodes at $(0,0,\pm k^\ast_z)$, where $k^\ast_z=\cos^{-1}(|m|/t_z)$ for $t_z>|m|$ and $m/t_0<1$. It supports localized one-dimensional \emph{hinge} modes for $|k_z| < k^\ast_z$ [see Fig.~\ref{Fig:HOTWeyl}(a)]. However, the corner localization of the hinge modes decreases monotonically as one approaches the Weyl nodes from the center of the Brillouin zone ($k_z=0$), similar to the situation with the Fermi arcs of a first-order Weyl semimetal (WSM)~\cite{roy-Fermiarc, arc-hinge}. Within this range of $k_z$, the quadrupolar moment is quantized to $0.5$, but vanishes for $|k_z|>k^\ast_z$ [see Fig.~\ref{Fig:HOTWeyl}(b)]. By contrast, for $\Delta_2=0$, four Weyl nodes appear at $(0,0,\pm k^\alpha_z)$, where $k^\alpha_z=\cos^{-1}([m+\alpha \Delta_1]/t_z)$ for $\alpha=\pm$. Four pairs of Weyl nodes can be found at $(\pm k_0, \pm k_0, \pm k^0_z)$ when $\Delta_1=0$, where $k_0=\sin^{-1}(\Delta_2/[\sqrt{2} t_0])$ and $k^0_z=\cos^{-1}([m-2t_0-2t_0 \cos(k_0)]/t_z)$. A complete analysis of three-dimensional second-order Weyl semimetals in the $(\Delta_1, \Delta_2)$ plane is left for a future investigation. It should be noted that so far only second-order Dirac semimetals (supporting linearly touching Kramers \emph{degenerate} valence and conduction bands) have been discussed in the literature~\cite{lin-hughes-DSM, wang1-2018, calugaru2019}, whereas we here demonstrate that it is conceivable to realize its Weyl counterparts (yielding linear touching between Kramers \emph{non-degenerate} bands), protected by an \emph{antiunitary symmetry}.

\emph{Summary and discussions}. To summarize, we identify an antiunitary operator ($A$) that assures the spectral symmetry of a two-dimensional HOT insulator [see Eq.~(\ref{Eq:HOTImodel})] and pins four corner modes at precise zero energy. Such an antiunitary symmetry allows us to construct a GHOTI for charged as well as neutral fermions, in terms of two additional perturbations [see Eq.~(\ref{Eq:genHOTIpert})], that continues to support corner localized zero-energy mode (see Figs.~\ref{Fig:PD_bandstructure} and \ref{Fig:GHOTI_CornerModes}), at least when they are small. In particular, our findings suggest that the corner localized Majorana zero modes of a HOT $p+id$ superconductor survive even in the presence of a weak Zeeman coupling and a parasitic or strain engineered $s$-wave pairing. Concomitantly, a transition between a HOT to trivial paired state can be triggered by tuning the strength of the external magnetic field or uniaxial strain, which can be instrumental for topological quantum computing based on Majorana fermions. The proposed anitiunitary symmetry protected corner and hinge modes can also be observed in highly tunable metamaterials, such as electrical circuits~\cite{junkai}.

\emph{Acknowledgments}.~The author thanks Vladimir Juri\v ci\' c, Soumya Bera, and Junkai Dong for discussions. B.R. was partially supported by the start-up grant from Lehigh University.


\begin{thebibliography}{}
\bibitem{antiunitary-comment} Alternatively, one can choose $\Gamma_1$ and $\Gamma_2$ ($\Gamma_3$ and $\Gamma_4$) to be purely imaginary (real), then $\Gamma_5$ is real and $A= \Gamma_5 K$.
\bibitem{antiunitary-wavefunction} If $|\Psi_E \rangle$ is an eigenstate of $\hat{h}^{\rm gen}_{\rm HOT}$ with energy $E$, then $A|\Psi_E \rangle \equiv |\Psi_{-E} \rangle$ is an also an eigenstate of $\hat{h}^{\rm gen}_{\rm HOT}$, but with energy $-E$, as $\{ \hat{h}^{\rm gen}_{\rm HOT}, A \}=0$. Therefore, the zero-energy manifold constituted by the corner modes with $E \approx 0$ (due to a finite system size) remains invariant under and protected by the antiunitary operator $A$.
\bibitem{arc-hinge} A first-order WSM can be constructed by stacking 2D quantum anomalous Hall insulators (QAHIs), occupying the $xy$ plane in the momentum space (along $k_z$, for example) between two Weyl nodes. The locus of the zero-energy modes associated with the 1D edge modes of each layer of QAHI constitutes the Fermi arc~\cite{roy-Fermiarc}. By contrast, a second-order WSM is constructed by stacking 2D GHOTI (supporting only pointlike corner modes) yielding 1D hinge states, as shown in Fig.~\ref{Fig:HOTWeyl}(a).
\end{thebibliography}
\end{document}